\documentclass[prb,aps,twocolumn,showpacs,bibnotes,superscriptaddress,epsf]{revtex4-1}

\usepackage{graphicx}
\usepackage{amsmath}
\usepackage{amssymb}
\usepackage{color}
\usepackage{dcolumn}
\usepackage{epsfig}
\usepackage{bm}
\usepackage{array}
\usepackage{multirow}
\usepackage[urlcolor=blue]{hyperref}
\usepackage{booktabs}
\usepackage{float}
\hypersetup{colorlinks=true, linkcolor=blue, citecolor=blue}

\begin{document}

\title{Lattice effect on the superexchange interaction in antiferromagnetic Bi$_2$Sr$_2$CaCu$_2$O$_8$}

\author{Jie Xin}
\affiliation{Department of Physics and State Key Laboratory of Surface Physics, Fudan University, Shanghai 200438, China}
\affiliation{Center for High Pressure Science and Technology Advanced Research, Shanghai 201203, China}

\author{Alexander G. Gavriliuk}
\affiliation{Institute for Nuclear Research, Russian Academy of Sciences, Moscow, Troitsk 108840, Russia}
\affiliation{FSRC Crystallography and Photonics of Russian Academy of Sciences, Moscow 119333, Russia}
\affiliation{REC Functional Nanomaterials, Immanuel Kant Baltic Federal University, Kaliningrad 236041, Russia}

\author{Jia-Wei Hu}
\affiliation{Center for High Pressure Science and Technology Advanced Research, Shanghai 201203, China}

\author{Jian-Bo Zhang}
\affiliation{Center for High Pressure Science and Technology Advanced Research, Shanghai 201203, China}

\author{Gen-Da Gu}
\affiliation{Condensed Matter Physics and Materials Science, Brookhaven National Laboratory, New York 11973, USA}

\author{Viktor V. Struzhkin}
\affiliation{Center for High Pressure Science and Technology Advanced Research, Shanghai 201203, China}

\author{Alexander F. Goncharov}
\affiliation{Earth and Planets Laboratory, Carnegie Institution for Science, Washington, DC 20015, USA}

\author{Hai-Qing Lin}
\affiliation{School of Physics, Zhejiang University, Hangzhou 310058, China}

\author{Ho-Kwang Mao}
\affiliation{Center for High Pressure Science and Technology Advanced Research, Shanghai 201203, China}

\author{Xiao-Jia Chen}
\email{xjchen2@gmail.com}
\affiliation{Center for High Pressure Science and Technology Advanced Research, Shanghai 201203, China}
\affiliation{School of Science, Harbin Institute of Technology, Shenzhen 518055, China}

\date{\today}

\begin{abstract}

The in-plane superexchange interaction $J$ of cuprate superconductors has long been suggested to be an important parameter for exploring their high-temperature superconductivity. The bilayer Bi$_2$Sr$_2$CaCu$_2$O$_{8+\delta}$ is the most studied system with high-quality single crystals in the wide doping range with the the same structure and phase. So far, the lattice parameter dependence of $J$ in its antiferromagnetic parent compound Bi$_2$Sr$_2$CaCu$_2$O$_{8}$ has not been established. By combining Raman scattering and x-ray diffraction techniques on the same sample in the same pressure environment, we obtain the evolution of both the two-magnon spectrum and the structural parameters with pressure up to nearly 30 GPa, The relationship between pressure or the in-plane lattice parameter and $J$ is thus established for Bi$_2$Sr$_2$CaCu$_2$O$_{8}$.  Over the studied pressure range, superconductivity does not appear in this parent compound based on a sensitive magnetic measurement technique. The effects of pressure and chemical doping on the superexchange interaction and structure and their implications for superconductivity are discussed from the comparison of the obtained experimental data with the existing experiments. The results and findings provide valuable information for the understanding of superconductivity and the future theory developments for superconductivity in cuprates. 

\end{abstract}

\maketitle

\section{Introduction}

Since the high-temperature superconductivity was discovered in cuprates\cite{bednorz,mkwu}, extensive experimental and theoretical researches have been carried out with many important findings. However, a comprehensive theoretical understanding of superconductivity in cuprates remains elusive. Resonating valence bonds\cite{ander,Anderson_2004} and spin-fluctuation mediated\cite{mont,scal} scenarios are among the most promising theoretical approaches in describing superconductivity in cuprates. In such spin-related theories, the Cu-O-Cu superexchange interaction, denoted by \textit{J}, is considered to be the cause of pairing. In the spin-fluctuation mediated scenario, electrons are paired up by exchanging magnetic spin fluctuations, which are also governed by \textit{J}\cite{scal}. The $J$ value can be determined conveniently from the two-magnon spectrum based on Raman scattering spectroscopy\cite{deve}. The quantitative determination of $J$ has been given for antiferromagnetic La$_{2}$CuO$_{4}$\cite{lyon,sugai3,singh} and YBa$_{2}$Cu$_{3}$O$_{6}$\cite{lyon2} based on the Raman scattering spectroscopy after the discoveries of superconductivity in these two families.  Extensive experiments further revealed that $J$ changes with doping in a similar manner as the pseudogap and pair peak in almost all cuprate superconductors\cite{sugai3,lyon2,sugai2,Sugaimagnon,zait,yli2}. These similarities indicate that the high-energy magnetic fluctuations are possibly indeed involved in the formation of the pseudogap and the Cooper pairing. Such an evolution from the high-energy excitation to superconducting quasiparticles has been suggested from coherent charge fluctuation spectroscopy\cite{mans}. The development from the two-magnon peak to a quasiparticle response with doping has been reproduced theoretically\cite{nlin}. Interestingly, there is a report for the existence of an approximately linear relationship between \textit{J} and the maximum critical temperature $T_{c}$ for many optimally doped cuprates\cite{Hg2022}. These results seemingly indicate the significance of \textit{J} and its connection to the superconductivity in cuprates. While emphasizing the importance of this parameter for the understanding of superconductivity, one will need to have the quantitative relationship between the lattice parameter and $J$. However, such information is rare\cite{PhysRevB.42.10785} and difficult to be drawn from the existing experiments of many superconducting families\cite{lyon,sugai3,singh,lyon2,sugai2,Sugaimagnon,zait,yli2,Hg2022}.   

Applying pressure has been proven to provide a clean and effective tool to tune the structural, magnetic, and superconducting properties\cite{mao}. Raman spectra of many materials including cuprates can be collected at high pressures\cite{gonch}. The high-pressure two-magnon spectra of the parent compounds\cite{aron,erem,struzh,maksimov} of two important cuprate families have been obtained by using Raman spectroscopy. High-pressure $J$ behaviors were thus established for La$_{2}$CuO$_{4}$\cite{aron}, Eu$_{2}$CuO$_{4}$\cite{erem,maksimov}, Sr$_{2}$CuCl$_{2}$O$_{2}$\cite{struzh}, and YBa$_{2}$Cu$_{3}$O$_{6.2}$\cite{maksimov}. To simplify the traditionally complicated theoretical expression for \textit{J}, the lattice parameter dependence of \textit{J} is expressed by a power law, $ J \sim d^{-n}$, where \textit{d} is the in-plane Cu-O-Cu bonding length and \textit{n} is an empirically determined constant. In order to establish the accurate relationship between $d$ and $J$, one needs to determine the structural evolution with pressure for the studied material at the same environments of composition, temperature, and pressure etc. Based on the structural information in the literature, the estimated relationships have been given for La$_2$CuO$_4$ with $J \sim d^{-(6.4\pm0.8)}$\cite{aron}, Eu$_{2}$CuO$_{4}$ and YBa$_2$Cu$_3$O$_{6.2}$ with $J \sim d^{-(3.0\pm0.5)}$\cite{maksimov}. Here the empirical constant $n$ does not differ too much with the one appeared in $J \sim d^{-(4\pm2)}$ for M$_2$CuO$_4$ (M=Pr, Nd, Sm, Eu, and Gd)\cite{PhysRevB.42.10785}. The well designed experiments on an insulating compound Bi$_{1.98}$Sr$_{2.06}$Y$_{0.68}$CaCu$_{2}$O$_{8.03}$ with neon as pressure transmitting medium gave $J \sim d^{-10.33}$\cite{cuk}. It remains unknown whether the large $n$ value is due to the lattice distortion or internal pressure induced by Y substitution for Ca in the studied system\cite{kaki}. So far, such an accurate relationship for an antiferromagnetic parent cuprate has not been established yet based on the simultaneous spectroscopy and diffraction measurements on the same sample in the same pressure environments. This is obviously an urgent subject in order to elucidate the spin-related pairing mechanism for superconductivity in high-$T_{c}$ cuprates.   

This work is designed to solve the above mentioned problem with the help of high-quality single crystals of Bi$_2$Sr$_2$CaCu$_2$O$_{8+\delta}$. The composition choice of the samples hopes to eliminate the possible external effect resulting from the elemental substitution and draw the information solely due to the application of pressure. After characterizing the antiferromagnetic feature of the chosen single crystal Bi$_2$Sr$_2$CaCu$_2$O$_{8}$ (with the doping level $\delta\sim0$), we perform high-pressure measurements of Raman scattering and x-ray diffraction (XRD) to obtain the spectroscopic and structural properties at high pressures. The evolution of the two-magnon spectrum with pressure for Bi$_2$Sr$_2$CaCu$_2$O$_{8}$ is obtained after the comparison of the spectroscopic character of a nearly optimally doped Bi$_2$Sr$_2$CaCu$_2$O$_{8+\delta}$ loaded together in the sample chamber near Bi$_2$Sr$_2$CaCu$_2$O$_{8}$. The pressure (or lattice parameter) dependence of $J$ is determined for Bi$_2$Sr$_2$CaCu$_2$O$_{8}$. Within the studied pressure regime and same pressure environment, the idea of pressure-induced superconductivity in Bi$_2$Sr$_2$CaCu$_2$O$_{8}$ is experimentally examined based on a highly sensitive magnetic detection technique. We also discuss the possible factors controlling superconductivity from the comparisons of the present and available experiments.    

\section{Experimental details}

The single crystals of Bi$_2$Sr$_2$CaCu$_2$O$_{8+\delta}$ grown using the floating zone technique were detailed previously\cite{gu}. Magnetization measurements were used to characterize the magnetic feature and superconducting transition of the synthesized samples by combining a superconducting quantum interference device magnetometer (Quantum Design MPMS3) and an in-house highly sensitive magnetic detection instrument. The latter was developed for detecting superconductivity for the sample with small dimensions in diamond anvil cell (DAC) at high pressures\cite{mao,sus,chen2010}. This technique is based on the suppression of the Meissner effect and quenching of superconductivity in the sample by an external magnetic field. The magnetic field applied to the sample inserted in the DAC mainly affects the signal coming from the sample. If the sample is superconducting, its transition temperature can be identified at the point where the signal goes to zero due to the disappearance of the Meissner effect. This highly sensitive technique has been used to discover superconductivity in compressed sulfur \cite{struzhkin1997} and lithium \cite{struzhkin2002} and study the pressure effect on \textit{T$_{c}$} in conventional superconductor NbN\cite{nbn} and various cuprates including  YBa$_2$Cu$_3$O$_{7-\delta}$\cite{pnas}, Bi$_{2}$Sr$_2$Ca$_{n-1}$Cu$_n$O$_{2n+4+\delta}$ ($n$=2,3) \cite{chen2004high,chen2010}, Tl$_{2}$Ba$_2$CaCu$_{2}$O$_{8+\delta}$ \cite{zhang2015effects}, and HgBa$_2$CuO$_{4+\delta}$\cite{Wang2014strain}. 

In the current study, we loaded nearly optimally doped Bi$_2$Sr$_2$CaCu$_2$O$_{8+\delta}$ crystal as a reference and an antiferromagnetic Bi$_2$Sr$_2$CaCu$_2$O$_{8}$ crystal separately into a DAC. The crystal was cut in the dimensions of approximately 100$\times$100$\times$20 $\mu$m$^{3}$. It was loaded with a ruby chip together into a sample chamber in a gasket made by non-magnetic Ni-Cr alloy. The gasket is pressed in between two diamond anvils with diameter of 300 $\mu$m in a DAC made by nonmagnetic Be-Cu alloy. The layout of such instruments and designs were given in detail previously\cite{mao,sus,chen2010}. The DAC was then mounted in a low-temperature cryostat with accessible optical windows, manufactured from Cryo Industries of America, lnc. The pressure can be mechanically tuned at demanded pressure and temperature. In such a design, Raman spectrum of the sample in DAC can be collected at the desired temperature and pressure. The temperature was monitored from the silicon diode sensor attached to the DAC in close proximity to the sample with a typical precision of $\pm$0.5 K. The pressure level was gauged based the shift of the $R$1 fluorescence line of ruby\cite{ruby} with the correction of temperature effect. For high-pressure magnetic susceptibility measurements, we changed and measured pressure at temperature of about 120 K. Below this temperature, the pressure can maintain a nearly constant value at low temperatures for the DAC and gasket in such an instrument.   

To have the consistency of the spectroscopy and structure data for the comparison, we conducted the measurements of high-pressure Raman scattering and XRD measurements on Bi$_2$Sr$_2$CaCu$_2$O$_{8}$ at room temperature. In fact, the spectroscopic signature of the two-magnon scattering for our studied Bi$_2$Sr$_2$CaCu$_2$O$_{8}$ is not strong. For comparison, we loaded this crystal together with a nearly optimally doped crystal nearby into a sample chamber (about 120 $\mu$m in diameter and 30 $\mu$m in thickness) of a DAC with a pair of diamond anvils in diameter of 300 $\mu$m. Both crystals were freshly cleaved with shining surfaces. A chip of ruby was put near the edge of the chamber filled with neon as the pressure medium. The Raman scattering signal was created by the Coherent sapphire laser device with 488 nm wavelength to illuminate an approximate 5$\times$5 $\mu$m$^{2}$ spot on the crystal surface. The low 2 mW power of the laser was chosen in order to avoid the over-heating of the sample. Raman spectra were collected in the back-scattering geometry. For each pressure of interest, we collected the Raman spectra of the two crystals one by one by keeping the same measurement conditions such as collecting time and laser grating. The Raman spectrum of the nearly optimally doped Bi$_2$Sr$_2$CaCu$_2$O$_{8+\delta}$ at each studied pressure exhibits a nearly flat background in the two-magnon scattering wavenumber range. The weak two-magnon scattering spectrum of Bi$_2$Sr$_2$CaCu$_2$O$_{8}$ at the same pressure can be picked up and enlarged by subtracting the nearly flat background from the nearly optimally doped Bi$_2$Sr$_2$CaCu$_2$O$_{8+\delta}$.   

Synchrotron XRD experiments were conducted at the 16BM-D beamline of the High Pressure Collaborative Access Team (HPCAT) at the Advanced Photon Source (APS) of Argonne National Laboratory, with an incident x-ray wavelength of 0.424603 {\AA}. Bi$_2$Sr$_2$CaCu$_2$O$_{8}$ crystal was finely ground into powders to obtain good diffraction data. The powders were pressed into a small piece. It was then loaded into a symmetric DAC with anvils in diameter of 300 $\mu$m together with a small ruby chip to monitor the pressure. Neon was used as the pressure-transmitting medium. The diffraction patterns were collected at room temperature at pressures up to 34 GPa. The two-dimensional patterns were integrated into 2$\theta$ vs. intensity data using Fit2D software\cite{hamm}. The diffraction patterns were analyzed using the Le Bail method in terms of the GSAS-\uppercase\expandafter{\romannumeral2} package\cite{toby}.

It should be mentioned that in all these measurements, the samples were kept in the neon environment. The experimental data for Bi$_2$Sr$_2$CaCu$_2$O$_{8}$ were collected and analyzed on the same single crystal. These considerations are expected to provide the consistent and reliable data of the studied system. 

\section{Results and discussion}

\subsection{Sample characterization}

\begin{figure}[tbp]
\includegraphics[width=0.4\textwidth]{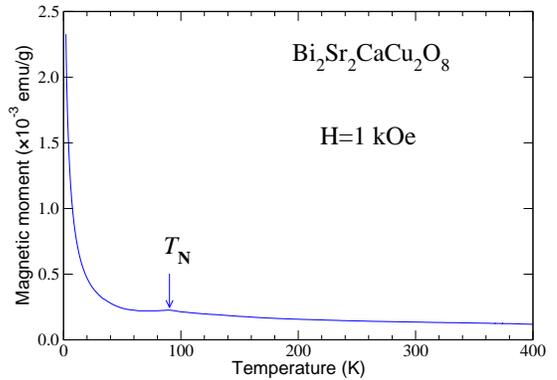}
\caption{Temperature dependence of the magnetic susceptibility of Bi$_2$Sr$_2$CaCu$_2$O$_8$ measured with a magnetic field of 1 kOe at ambient pressure.}
\label{mag}
\end{figure}

The characterization of the quantum ground state of Bi$_2$Sr$_2$CaCu$_2$O$_8$ is given from the magnetization measurements (Fig. \ref{mag}). There is an antifferomagntic-paramagnetic transition at the N\'eel temperature (\textit{T$_N$}) of 90 K, indicated by the kink in the temperature dependence of the magnetic moment in a magnetic field directed along the \textit{c} axis. No evidence for supporting superconductivity is provided from the magnetic susceptibility data down to 2 K. The kink for the antiferromagnetic transition in studied Bi$_2$Sr$_2$CaCu$_2$O$_8$ is similar to those found in other antiferromagnetic cuprates such as Bi$_2$Sr$_2$Ca$_{1-x}$Y$_x$Cu$_2$O$_{8+\delta}$\cite{Fukushima_1988}, La$_2$CuO$_{4+\delta}$\cite{BELEVTSEV20101307} and Bi$_2$SrLaCuO$_y$\cite{PhysRevB.41.6418}. No magnetic phase transition has ever been reported for Bi$_2$Sr$_2$CaCu$_2$O$_{8+\delta}$ \cite{PhysRevB.41.6418}. The \textit{T$_N$} value obtained for the studied Bi$_2$Sr$_2$CaCu$_2$O$_8$ compound is 20 K higher than those in the Bi$_2$Sr$_2$Ca$_{1-x}$Y$_x$Cu$_2$O$_{8+\delta}$ system\cite{Fukushima_1988}. This result indicates the good antiferromagnetic character of our sample. Its relatively high \textit{T$_N$} value also corresponds to the extremely low content of excess oxygen. Thus, the content of excess oxygen in the studied sample is assumed to be zero (that is, $\delta\sim0$). Therefore, we have a sample in the normal composition of Bi$_2$Sr$_2$CaCu$_2$O$_8$ with the antiferromagnetic nature for further investigations.

\subsection{Two-magnon scattering at high pressures}

Antiferromagnetism is the striking physical feature of the parent cuprates. A classical spin-wave is acoustic-like, which means that a magnon (single-spin flip) excitation can hardly be observed in the inelastic light scattering channel. However, two neighboring spins on Cu sites bridged by intermediate oxygen ions may reverse their spin directions simultaneously. This process meets the requirements of selection rules for the so-called two-magnon process permitted in Raman scattering. The double spin-flop excitations (two-magnon) on CuO$_2$ layers could be directly probed by the Raman scattering technique. As the simultaneous reversal of two adjacent spins costs the energy of the nearest six superexchange coupling constants, the two-magnon peak should appear at about 3\textit{J} in energy. Due to the magnon-magnon interaction, the ratio slightly deviates from 3, and  this ratio is taken as 2.7 based on the two-dimensional Heisenberg model\cite{weber}. 

\begin{figure}[tbp]
\includegraphics[width=0.45\textwidth]{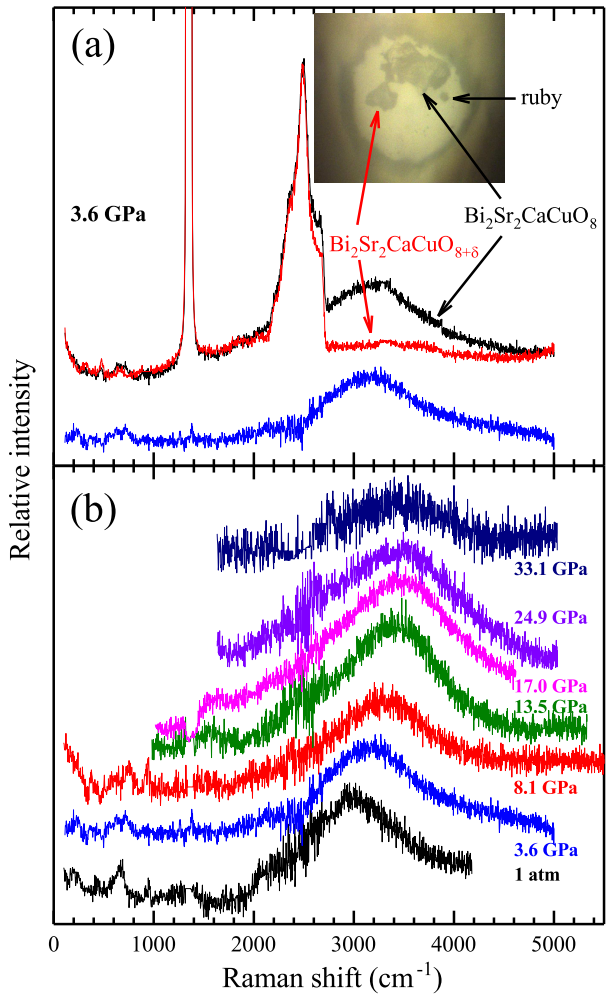}
\caption{Two-magnon Raman scattering spectra for antiferromagnetic Bi$_2$Sr$_2$CaCu$_2$O$_8$ under pressure. (a) Raman spectra of the antiferromagnetic (black curve) and nearly optimally doped (red curve) Bi$_2$Sr$_2$CaCu$_2$O$_{8+\delta}$ at pressure of 3.6 GPa. The blue curve corresponds to the spectrum subtraction from the signal of the antiferromagnetic sample to that of the nearly optimally doped one. Inset: Photograph of two crystals in neon environment with a ruby chip. (b) Subtracted two-magnon scattering spectra of Bi$_2$Sr$_2$CaCu$_2$O$_8$ at representative pressures in the compression run.}
\label{AFM-2M-1}
\end{figure}

For a parent cuprate, previous Raman scattering studies detected the two-magnon peak around 3000 cm$^{-1}$ at ambient pressure, corresponding to a $J$ of $\sim$1000 cm$^{-1}$ (about 125 meV)\cite{lyon,sugai3,singh,lyon2}. This spectrum range just overlaps with the two-phonon Raman peak of the diamond (around 2500 cm$^{-1}$). Therefore, the collection of the two-magnon spectrum is a challenge by using diamond anvils. This is probably a reason why the high-pressure behaviors of the two-magnon Raman scattering for many parent cuprates have not been reached yet. To solve this critical problem, we loaded the antiferromagnetic Bi$_2$Sr$_2$CaCu$_2$O$_8$ crystal together with a nearly optimally doped Bi$_2$Sr$_2$CaCu$_2$O$_{8+\delta}$ into a DAC (Inset of Fig. \ref{AFM-2M-1}(a)). By performing the measurements at the same conditions, we obtained the Raman spectra of them. As an example, the Raman spectra at 3.6 GPa are shown in Fig. \ref{AFM-2M-1}(a). With the assumption of the same active Raman feature from the used diamond anvils or medium for the studied two crystals at the messured pressure, the strong two-magnon scattering spectrum of Bi$_2$Sr$_2$CaCu$_2$O$_8$ can be derived by subtracting the collected spectrum from that of Bi$_2$Sr$_2$CaCu$_2$O$_{8+\delta}$. This procedure turned out to be very successful. As can be seen, the two-magnon frequency of Bi$_2$Sr$_2$CaCu$_2$O$_8$ at ambient pressure is well situated in the trend of Bi$_2$Sr$_2$CaCu$_2$O$_{8+\delta}$ with the different dopants\cite{sugai2,Sugaimagnon}. The positions of the phonon modes at the low wavenumbers are also consistent with those reported from Raman measurements\cite{PhysRevLett.82.3524,PhysRevB.69.054511}.

\begin{figure}[tbp]
\includegraphics[width=0.50\textwidth]{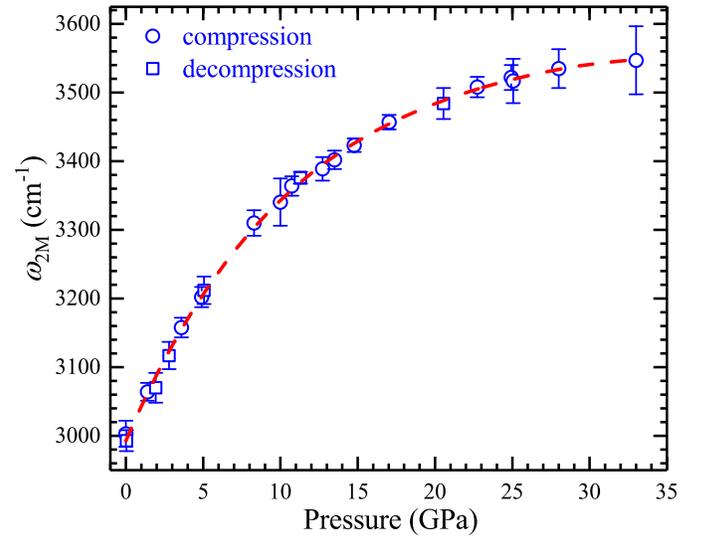}
\caption{Pressure dependence of the two-magnon frequency in antiferromagnetic Bi$_2$Sr$_2$CaCu$_2$O$_8$ in the compression and decompression run. The dished curve is the polynomial fitting to the data points for the guide to the eyes.}
\label{AFM-2M-2}
\end{figure}

\begin{figure*}[tbp]
\includegraphics[width=0.9\textwidth]{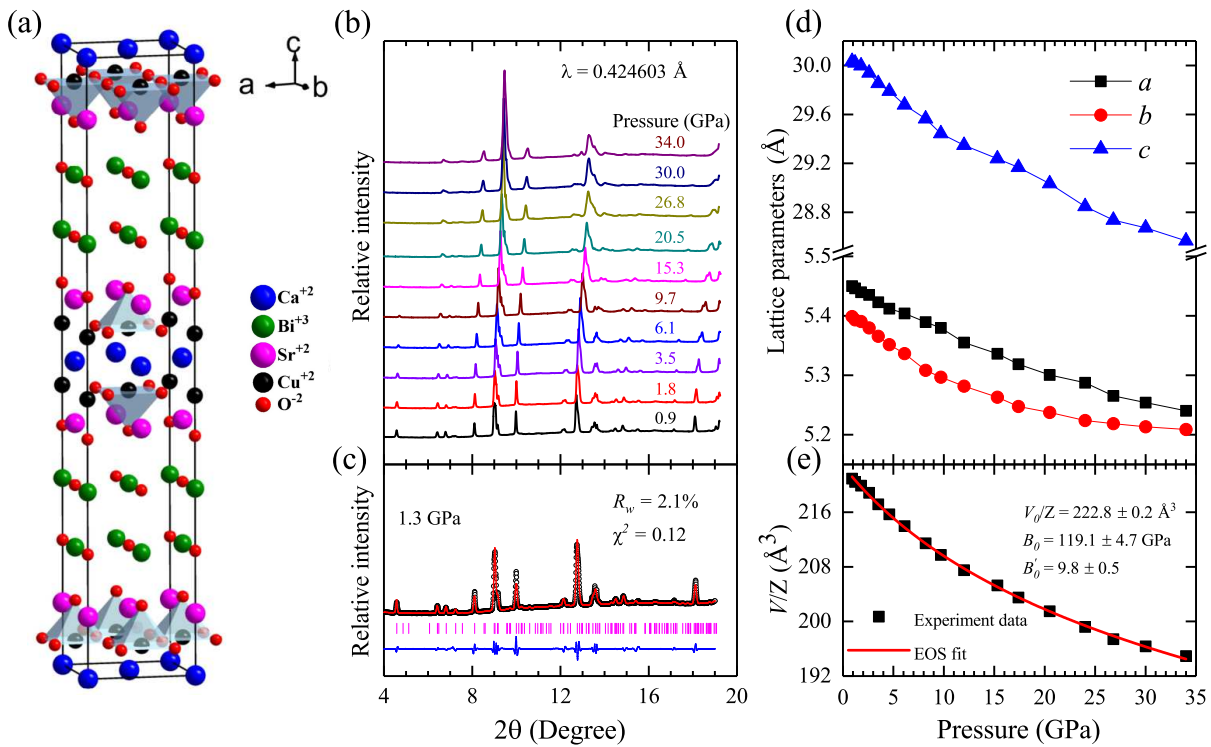}
\caption{Structural properties of antiferromagnetic Bi$_2$Sr$_2$CaCu$_2$O$_8$ at high pressures. (a) The schematic orthorhombic structure with the space group of $Amaa$. (b) The integrated x-ray diffraction data at representative pressures. (c) The observed XRD patterns (circles), Le Bail fitting (red line), and their difference (blue line) between the observed and calculated profiles at pressure of 1.3 GPa. The vertical bars refer to the peak positions. (d) The lattice parameters along the \textit{a}, \textit{b}, and \textit{c} axis as a function of pressure. (e) Pressure dependence of the unit-cell volume $V/Z$.}
\label{AFM-XRD}
\end{figure*}

Figure \ref{AFM-2M-1}(b) shows the two-magnon Raman scattering spectra of Bi$_2$Sr$_2$CaCu$_2$O$_8$ at various pressures up to 33.1 GPa. The two-magnon scattering signal was recorded during the compression and decompression run. As pressure is increased, the two-magnon peak broadens and shifts to higher wavenumbers (or energies) with the decrease of the intensity. These behaviors are consistent with the early reports for La$_2$CuO$_4$ up to 10 GPa\cite{aron}, Eu$_{2}$CuO$_{4}$ up to 41 GPa\cite{maksimov}, YBa$_2$Cu$_3$O$_{6.2}$ up to 43 GPa\cite{maksimov}, and Bi$_{1.98}$Sr$_{2.06}$Y$_{0.68}$CaCu$_{2}$O$_{8.03}$ up to 32 GPa\cite{cuk}. The evolution of the two-magnon frequency ($\omega_{2M}$) with pressure for Bi$_2$Sr$_2$CaCu$_2$O$_8$ is given in Fig. \ref{AFM-2M-2}. As can be seen, $\omega_{2M}$ increases continuously from nearly 3000 to 3550 cm$^{-1}$ with increasing pressure within the studied pressure range, as observed previously for La$_2$CuO$_4$\cite{aron}, Sr$_{2}$CuCl$_{2}$O$_{2}$\cite{struzh}, and  Bi$_{1.98}$Sr$_{2.06}$Y$_{0.68}$CaCu$_{2}$O$_{8.03}$\cite{cuk}. It is interesting to notice the nearly linear change of $\omega_{2M}$ with pressure were observed in the early studies on Eu$_{2}$CuO$_{4}$\cite{maksimov} and YBa$_2$Cu$_3$O$_{6.2}$\cite{maksimov} over the studied pressure range. The increase of $\omega_{2M}$ for Bi$_2$Sr$_2$CaCu$_2$O$_8$ upon the applied pressure to 33 GPa is almost the same as La$_2$CuO$_4$\cite{aron} within 10 GPa. However, Bi$_{1.98}$Sr$_{2.06}$Y$_{0.68}$CaCu$_{2}$O$_{8.03}$\cite{cuk} exhibits a large increase of $\omega_{2M}$ from the nearly same initial value to above 4000 cm$^{-1}$ within the nearly same pressure range as the present study. In all these measurements on cuprates, an increasing $\omega_{2M}$ upon lattice compression is generic. This can be considered as the common feature of $\omega_{2M}$ for cuprates at high pressures.  

Superconductivity in cuprates was found to develop with the suppression of the antiferromagnetic order upon doping. It has been generally believed that the application of pressure has the equivalent effect on superconductivity as the chemical doping in these compounds. For comparison, the two-magnon peak behaves with doping in an opposite way as pressure.  Upon doping, the two-magnon peak indeed becomes broader but moves to lower wavenumber\cite{sugai3,lyon2,sugai2,Sugaimagnon,zait,yli2,rubh,blum2}. The totally different change directions for the two-magnon frequency from doping and pressure suggest their different effects if the superexchange interaction is tightly related to superconductivity.

\subsection{Structural evolution with pressure}

To obtain the evolution of $\omega_{2M}$ (thus \textit{J}) against the lattice parameters, we conducted synchrotron XRD measurements for Bi$_2$Sr$_2$CaCu$_2$O$_8$ at various pressures up to 34 GPa. The schematic crystal structure of this material, belonging to the space group of $Amaa$\cite{gao}, is shown in Fig. \ref{AFM-XRD}(a). This structure consists of two equivalent perovskite sub-cells with respect to each other along the diagonal direction in the giving \textit{a}-\textit{b} plane and bonded by Van der Waals force. Each sub-cell consists of two insulating BiO layers and two SrO layers. In the middle of the sub-cell, two structurally equivalent CuO$_2$ layers are separated by the Ca layer. These layers are stacked in the sequence of BiO-SrO-CuO$_2$-Ca-CuO$_2$-SrO-BiO.

The powder XRD results of Bi$_2$Sr$_2$CaCu$_2$O$_8$ at high pressures and room temperature are presented in Fig. \ref{AFM-XRD}(b). With the application of pressure, all diffraction peaks gradually shift to high angles, indicating the lattice shrinking. No new diffraction peaks appear. There is also no peak merging or splitting. These observations indicate no structural transformation under pressure. The compound maintains the same structure at the studied pressure range, consistent with previous high-pressure structural studies on this system\cite{Olsen_1991,Zhang_2013}. 

Different from other cuprate families, Bi-based system possesses an incommensurate structure modulation along the \textit{b} axis\cite{PhysRevB.38.893,gao1988incommensurate,petricek1990x,slezak2008imaging}. The evolution of the modulation vector for bilayer Bi$_2$Sr$_2$CaCu$_2$O$_{8+\delta}$ with pressure has never been reported. We observed a weak but regular peak in low angle of about 1$^{o}$ at each applied pressure (not shown in Fig. \ref{AFM-XRD}(b)), consistent with the early observations\cite{petricek1990x,slezak2008imaging}. This peak is considered the first-order diffraction peak from the modulation superstructure. The component of modulation vector $\vec{b}^*$ can be calculated from this peak position. The vector component $b^*$ is found to be around 0.22 and remains unchanged within the studied pressure range.

The refinement of the crystal structure of Bi$_2$Sr$_2$CaCu$_2$O$_8$ is not an easy task. We believe that the lattice parameters can be determined by Le Bail fitting to the average structure model with sufficient accuracy. Figure \ref{AFM-XRD}(c) illustrates the powder XRD data and Le Bail fitting for Bi$_2$Sr$_2$CaCu$_2$O$_8$ at pressure of 1.3 GPa based on the structure model depicted in Fig. \ref{AFM-XRD}(a). The fitting provides reasonable factors of \textit{R$_w$} = 2.1\% and $\chi$$^2$ = 0.12. The similar fittings to the collected diffraction patterns provide the lattice parameters with comparable quality up to 34 GPa. These results indicate that Bi$_2$Sr$_2$CaCu$_2$O$_8$ keeps in an orthorhombic structure within the studied pressure range. Figure \ref{AFM-XRD}(d) shows the pressure dependence of the lattice parameters along the \textit{a}, \textit{b}, and \textit{c} axis. The corresponding unit-cell volume \textit{V} with \textit{Z} of 4 being the number of the chemical formula per unit cell is presented in Fig. \ref{AFM-XRD}(e). The lattice parameters and cell volume decrease gradually with the increase of pressure. The pressure-volume data are fitted using the third-order Birch-Murnaghan equation of states (EOS)\cite{PhysRev.71.809}, as shown by the curve in Fig. \ref{AFM-XRD}(e). The fitting gives the bulk modulus at zero pressure \textit{B$_0$} = 119.1$\pm$4.7 GPa, the initial unit cell volume per chemical formula \textit{V$_0$/Z} = 222.8$\pm$0.2 {\AA}$^3$ and the first pressure derivative of the bulk modulus at zero pressure \textit{B$_0'$} = 9.8$\pm$0.5 for Bi$_2$Sr$_2$CaCu$_2$O$_8$. The obtained \textit{B$_0$} value is bigger than the one reported for underdoped Bi$_2$Sr$_2$CaCu$_2$O$_{8+\delta}$ with \textit{B$_0$} = 114.9$\pm$6.4 GPa\cite{Zhang_2013}, and much larger than the reported value of 73 GPa\cite{TAJIMA1989237}, 62.5 GPa\cite{Olsen_1991}, and 68.6 GPa\cite{JIANDINGYU199745} for Bi-based compounds. The dramatic difference in \textit{B$_0$} comes from the difference in samples and the experimental conditions.

The change of the obtained lattice parameters of antiferromagnetic Bi$_2$Sr$_2$CaCu$_2$O$_8$ under pressure is again used to compare the doping effect on the crystal structure. Amongst all discovered cuprate superconductors, bilayer Bi$_2$Sr$_2$CaCu$_2$O$_{8+\delta}$ is a most studied superconducting system with the complete phase diagram\cite{sieb,droz} over the whole doping range covering from the slightly underdoped region through the optimal level to the heavily overdoped regime. Interestingly, the structural properties of this system as a function of oxygen content over the whole doping range have not fully been investigated yet. To date, no systematic evolution of the lattice parameter(s) with doping is available for Bi$_2$Sr$_2$CaCu$_2$O$_{8+\delta}$ over a relatively wide doping range\cite{groen,liang}, probably due to the complexity in the structure determination\cite{PhysRevB.38.893,gao1988incommensurate,petricek1990x,slezak2008imaging}. The annealing of as-grown samples with increasing partial oxygen pressure leads to the increase of the oxygen content. At such conditions, $T_c$ was reported to decrease with oxygen pressure in a similar way as the reduction of the $c$-axis lattice constant\cite{groen}. This observation seemingly indicates the $c$-axis lattice parameter decreases in the overdoped regime. On the other hand, the similar trend for the $c$-axis lattice parameter was found with the loss of oxygen for the single-phase Bi$_2$Sr$_2$CaCu$_2$O$_{8+\delta}$\cite{liang}. Since the annealing of the as-grown crystal simulates the underdoping effect, this behavior suggests the reduction of the $c$-axis lattice parameter in the underdoped regime. Although the exact $c$ values do not match each other even for their different as-grown samples, these two sets of experiments\cite{groen,liang} indicate the continuous decrease of the $c$-axis lattice parameter upon the carrier concentration. There are not any clear trends for the lattice parameters along the $a$ and $b$ direction from these experiments\cite{groen,liang}. The observed reduction of the $c$ lattice parameter for Bi$_2$Sr$_2$CaCu$_2$O$_8$ may imply the increase of the carrier concentration by the application of pressure. 

\begin{figure}[tbp]
\includegraphics[width=0.45\textwidth]{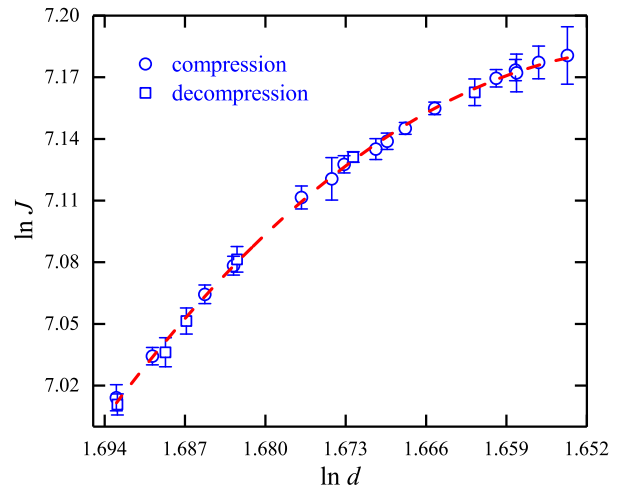}
\caption{The double logarithmic plot of in-plane \textit{J} versus the average in-plane lattice parameter \textit{d}. The dashed curve is the fitting to the data points.}
\label{AFM-2M-3}
\end{figure}

\subsection{Relation between the in-plane superexchange interaction \textit{J} and the lattice parameter}

Having determined the pressure dependence of $\omega_{2M}$ and the equation of states for Bi$_2$Sr$_2$CaCu$_2$O$_8$, we can obtain the lattice parameter dependence of $\omega_{2M}$ after converting the pressure into the lattice parameter. Such a dependence for the in-plane superexchange interaction \textit{J} can be directly obtained through the expression $\omega_{2M} = 2.7\textit{J}$\cite{weber}. For layered cuprates, the in-plane superexchange interaction \textit{J} is insensitive to the out-of-plane coupling. One can only consider the parameters within the CuO$_2$ plane. There is a slight difference between the lattice parameters along the \textit{a} and \textit{b} axis for Bi$_2$Sr$_2$CaCu$_2$O$_8$. It is reasonable to define the average lattice constant in the \textit{a-b} plane as $d = \sqrt{ab}$. The \textit{d} value at the given pressure for the corresponding $\omega_{2M}$ (or $J$) is thus obtained from the determined equation of states based on the expression of $V/Z=d^{2}$$c$/4.

To determine the exponent \textit{n} in $J \sim d^{-n}$ for Bi$_2$Sr$_2$CaCu$_2$O$_8$, we plotted $\ln J$ as a function of $\ln d$ in Fig. \ref{AFM-2M-3}. We observed a linear dependence of $\ln J$ on $\ln d$ within 6 GPa. By performing a linear fitting of the $\ln J$ - $\ln d$ data in this pressure range, we obtained the relationship of $J \sim d^{-(6.6\pm0.2)}$. At high pressures ($\textgreater$ 6 GPa), the dependence of $\ln J$ on $\ln d$ gradually deviates downwards from linear behavior, and the deviation increases with pressure. This indicates that the exponent \textit{n} in the empirical formula $J \sim d^{-n}$ decreases with increasing pressure, implying reduced sensitivity of \textit{J} to \textit{d} at higher pressures as observed in La$_2$CuO$_4$\cite{aron}.

Compared with other cuprates, the lattice parameter dependence of $J$ in Bi$_2$Sr$_2$CaCu$_2$O$_8$ is close to those of La$_2$CuO$_4$ ($J \sim d^{-(6.4\pm0.8)}$)\cite{aron}, Eu$_{2}$CuO$_{4}$ and YBa$_2$Cu$_3$O$_{6.2}$ ($J \sim d^{-(3.0\pm0.5)}$)\cite{maksimov}, and M$_2$CuO$_4$ (M = Pr, Nd, Sm, Eu and Gd) ($J \sim d^{-(4\pm2)}$) \cite{PhysRevB.42.10785}, but differs substantially from that of Bi$_{1.98}$Sr$_{2.06}$Y$_{0.68}$CaCu$_{2}$O$_{8+\delta}$ ($J \sim d^{-10.33}$)\cite{cuk}. Considering the fact that this relationship is established based on the simultaneous Raman and XRD measurements on the same characterized crystal at the nearly equal experimental conditions, we hope it to shed light on important features for the establishment and understanding of the mechanism of superconductivity in high-$T_c$ cuprates. 

\subsection{Examination of possible superconductivity under pressure}

\begin{figure}[tbp]
\includegraphics[width=0.5\textwidth]{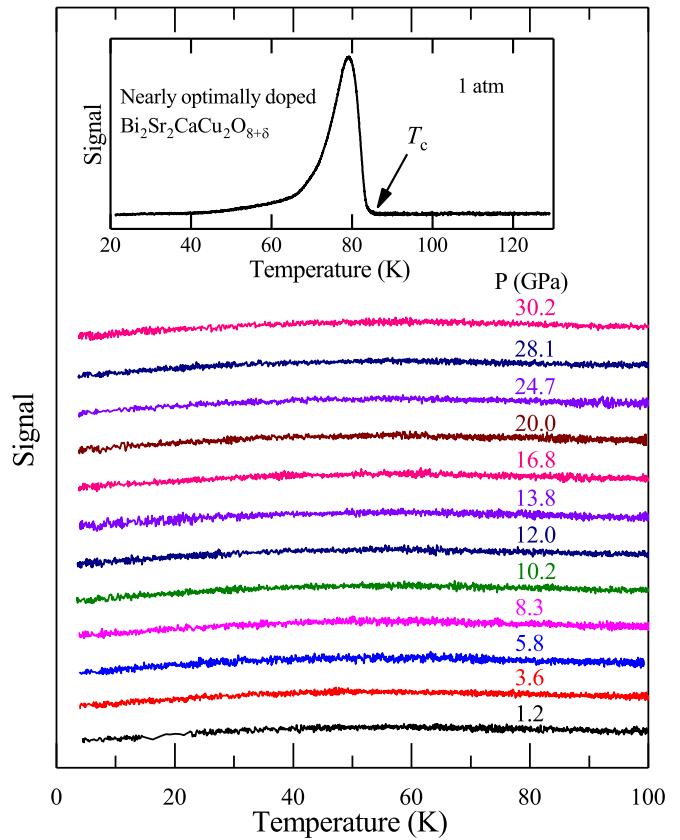}
\caption{Magnetic susceptibility signal versus temperature for antiferromagnetic Bi$_2$Sr$_2$CaCu$_2$O$_8$ single crystal at applied pressures up to 30.2 GPa. Inset: Magnetic susceptibility signal for nearly optimal doped Bi$_2$Sr$_2$CaCu$_2$O$_{8+\delta}$ at ambient pressure.}
\label{AFM-Mag}
\end{figure}

It has been generally believed that superconductivity can be created by doping or pressure in high-$T_c$ cuprates. This belief has been proven from the similar phase diagram of $T_{c}$ with these two tuning modes in strongly correlated superconductors such as heavy fermions\cite{math,park,norm} or iron-based compounds\cite{rott,alir,kimb}. In those superconductors, the structure shows the same behaviors of the lattice parameters with pressure and chemical substitution\cite{kimb}. For antiferromagnetic Bi$_2$Sr$_2$CaCu$_2$O$_8$, the shrinking of the lattice under pressure (Fig. \ref{AFM-XRD}) was found in a way similar to the effect of doping\cite{groen,liang}.  In fact, the reduction of the lattice parameters with the application of pressure is far beyond those found with doping\cite{groen,liang}. It is reasonable to examine whether superconductivity could be induced in Bi$_2$Sr$_2$CaCu$_2$O$_8$ just like other strongly correlated systems. We thus performed the magnetic susceptibility measurements on Bi$_2$Sr$_2$CaCu$_2$O$_8$ crystal with the same pressure environment as the Raman scattering and XRD measurements in the almost same pressure range. The results for such measurements at high pressures up to 30.2 GPa are presented  in Fig. \ref{AFM-Mag}. The pressure was applied and measured at a temperature of about 120 K, below which the pressure was kept near constant. For each pressure, the magnetic susceptibility data were collected in the warming run. For the comparison, we show the temperature dependence of the collected signal for nearly optimally doped Bi$_2$Sr$_2$CaCu$_2$O$_{8+\delta}$ at ambient pressure in the inset of Fig. \ref{AFM-Mag}. In both experiments, the crystal was cut in the nearly same size for the measurements. 

The magnetic susceptibility signal of nearly optimally doped Bi$_2$Sr$_2$CaCu$_2$O$_{8+\delta}$ exhibits a sudden jump to the peak from the flat background at temperature of 86 K followed by the gradual decline to lower temperatures (inset of Fig. \ref{AFM-Mag}). The starting jump point from the high-temperature side is defined as the superconducting transition due to the appearance of the Meissner effect when entering the superconducting state. Such a character for superconductivity has been observed in previous studies  \cite{struzhkin1997,struzhkin2002,nbn,chen2004high,chen2010,zhang2015effects,Wang2014strain} based on the developed high-pressure magnetic measurement technique\cite{sus}.

The almost identical and featureless signals were found for Bi$_2$Sr$_2$CaCu$_2$O$_8$ at various pressures (Fig. \ref{AFM-Mag}). This behavior is substantially different from the reference signal for the nearly optimally doped Bi$_2$Sr$_2$CaCu$_2$O$_{8+\delta}$. The typical character for superconductivity can not be detected in the measured range of temperature and pressure. These observations demonstrate that unlike heavy fermions\cite{math} or iron-based compounds\cite{alir}, Bi$_2$Sr$_2$CaCu$_2$O$_8$ does not exhibit superconductivity by the applied pressures up to 30 GPa. 

\subsection{What parameters matter for superconductivity}

The absence of superconductivity in antiferromagnetic sample upon lattice compression with pressure provides an excellent platform to test what parameters matter for superconductivity. As can be seen from Fig. \ref{AFM-XRD}, the application of pressure substantially shrinks the unit cell of Bi$_2$Sr$_2$CaCu$_2$O$_8$ compared to the effect due to doping\cite{groen,liang}. This indicates that doping alone is not enough for inducing superconductivity. To drive a system to superconduct, the pair formation is essential. For high-$T_{c}$ cuprates, the decrease of $J$ was generally observed with doping\cite{sugai3,lyon2,sugai2,Sugaimagnon,zait,yli2,rubh,blum2}. In such a way, quasiparticles may gain the energy for pairing with the expense of $J$\cite{nlin} if spin-mediated pairing works. The observed increase of $J$ in Bi$_2$Sr$_2$CaCu$_2$O$_8$ with pressure (Fig. \ref{AFM-2M-2}) indicates that the antiferromagnetic order does not collapse in the studied pressure range. Once the antiferromagnetism can be completely suppressed in this compound, the introduction of superconductivity is possible.

The reported nearly constant pressure derivative of $T_{c}$ ($dT_{c}$/$dP$)\cite{sieb} and universal parabolic evolution of $T_{c}$ with pressure\cite{chen2004high} for Bi$_2$Sr$_2$CaCu$_2$O$_{8+\delta}$ with different oxygen contents strongly suggest that another independent parameter controls the superconductivity besides the carrier concentration due to the chemical doping or elemental substitution. The unchanged magnetic excitations in La$_{2-x}$Sr$_{x}$CuO$_{4}$ across the entire doping regime of the phase diagram\cite{dean} indicate that $T_{c}$ could be controlled by other factors rather than the magnetic excitations themselves. An inelastic neutron scattering study on La$_{2-x}$Sr$_{x}$CuO$_{4}$ revealed a strong coupling between the spin excitations and optical phonons\cite{ikeu}. Such a coupling might be important for the formation of the charge-density wave states and superconductivity in these materials\cite{struzhkin2016}. The quantitative spectroscopic analysis\cite{shen} of one-dimensional antiferromagnetic cuprate chains suggested that the observed anomalously strong near-neighbor attraction probably originates from electron-phonon coupling. Indeed, without the consideration of electron-phonon interactions, it is hard to explain the observed sizable oxygen isotope effect and its evolution with adding CuO$_{2}$ plane in Bi$_2$Sr$_2$Ca$_{n-1}$Cu$_{n}$O$_{2n+4+\delta}$ ($n$=1,2,3)\cite{iso}. The electronic and phononic contributions to the pairing strengths can be distinguished from the time resolution spectroscopy measurements due to their different lifetimes\cite{conte,chia}. It was found that both electronic excitations and electron-phonon interactions can not be neglected\cite{conte,chia}. For Bi$_2$Sr$_2$CaCu$_2$O$_{8+\delta}$, the electron-phonon coupling strength was found to behave with doping in a way similar to the $T_{c}$ behavior, while the electron-spin fluctuation coupling strength decreases monotonically with doping\cite{chia}. Therefore, another parameter independent of the carrier concentration is the pairing strength itself. If the strong interactions can be induced upon lattice compression with the suppression of the antiferromagnetic order, the superconductivity in Bi$_2$Sr$_2$CaCu$_2$O$_8$ can be expected. In this aspect, the observed signature of superconductivity\cite{cuk2010} in Bi$_{1.98}$Sr$_{2.06}$Y$_{0.68}$CaCu$_{2}$O$_{8+\delta}$ at modest pressure of a few GPa is likely due to the interaction enhancement through the lattice modulation caused by Y substitution. 

Note that the doping dependence of the electron-spin fluctuation coupling strength\cite{chia} in Bi$_2$Sr$_2$CaCu$_2$O$_{8+\delta}$ is similar to the evolution of $J$ with doping\cite{sugai2,Sugaimagnon}. This similarity indicates that spin fluctuations or electronic excitations need to develop at the expense of the antiferromagnetic interactions. However, the electron-phonon coupling strength is larger than the electron-spin fluctuation coupling strength over the large doping region of the superconducting phase diagram\cite{chia} . Interestingly, the electron-phonon coupling strength varies parabolically with doping\cite{chia}, similar to the observed behaviors of the energy gap $\Delta$ and $2\Delta/k_{B}T_{c}$ as already reported from tunneling measurements from the underdoped regime\cite{miya,miya2} to the overdoped one\cite{dewi}.  

\section{Conclusions}

In conclusion, we have conducted a comprehensive study on antiferromagnetic Bi$_2$Sr$_2$CaCu$_2$O$_8$ under high pressures, using Raman scattering, X-ray diffraction, and magnetic susceptibility measurements. Raman scattering measurements revealed an increase in the two-magnon frequency from approximately 3000 cm$^{-1}$ at ambient pressure to 3550 cm$^{-1}$ at 33.1 GPa within the applied pressure range. The lattice parameters were observed to decrease monotonically with increasing pressure without any indication for the structural transition within the studied pressure range. Under compression, the in-plane superexchange interaction \textit{J} was found to increase with pressure and follow $J \sim d^{-(6.6\pm0.2)}$ with the in-place spacing distance at low pressures. The effects of pressure and chemical doping on the superexchange interaction and structure were discussed and compared. The implications of these parameters on superconductivity were given. Over the studied pressure range, we did not detect pressure-induced superconductivity in antiferromagnetic Bi$_2$Sr$_2$CaCu$_2$O$_8$. Possible factors for controlling superconductivity in high-$T_{c}$ cuprates were thus drawn from the comparisons of the experiments.

\section{acknowledgements}

The work at HPSTAR was supported by the National Key R$\&$D Program of China (Grant No. 2018YFA0305900). The work at HIT was supported from the Shenzhen Science and Technology Program (Grant No. KQTD20200820113045081) and the Basic Research Program of Shenzhen (Grant No. JCYJ20200109112810241). G.D.G was supported by U.S. Department of Energy (DOE) (DE-SC0012704). V.V.S. and H.K.M. acknowledge the financial support from Shanghai Science and Technology Committee, China (No. 22JC1410300) and Shanghai Key Laboratory of Materials Frontier Research in Extreme Environments, China (No. 22dz2260800). The XRD measurements were performed at HPCAT (Sector 16), Advanced Photon Source (APS), Argonne National Laboratory. HPCAT operations are supported by DOE-NNSA's Office of Experimental Sciences. The Advanced Photon Source is a U.S. DOE Office of Science User Facility operated for the DOE Office of Science by Argonne National Laboratory under Contract No. DE-AC02-06CH11357.

\end{document}